\documentclass[pdflatex,sn-mathphys-num]{sn-jnl}

\usepackage{graphicx}%
\usepackage{multirow}%
\usepackage{amsmath,amssymb,amsfonts}%
\usepackage{amsthm}%
\usepackage{mathrsfs}%
\usepackage[title]{appendix}%
\usepackage{xcolor}%
\usepackage{textcomp}%
\usepackage{manyfoot}%
\usepackage{booktabs}%
\usepackage{algorithm}%
\usepackage{algorithmicx}%
\usepackage{algpseudocode}%
\usepackage{listings}%
\usepackage{pdfpages,multirow,ragged2e,ulem} %
\usepackage{physics}
\usepackage{physunits}
\usepackage[utf8]{inputenc} 

\theoremstyle{thmstyleone}%
\theoremstyle{thmstyletwo}%

\theoremstyle{thmstylethree}%

\begin{document}

\title[Article Title]{Laboratory Tests of Laser Control of Electron Beams for Future Colliders}


\author[1]{\fnm{C.} \sur{Munting}}

\author[1]{\fnm{P.} \sur{Kicsiny}}

\author[1]{\fnm{E.} \sur{Barbi}}

\author[1]{\fnm{N.} \sur{Gonzalez}}

\author*[1]{\fnm{S.} \sur{Gessner}}\email{sgess@slac.stanford.edu}

\author[2]{\fnm{I.} \sur{Drebot}}

\affil[1]{\orgname{SLAC National Accelerator Laboratory}, \orgaddress{\street{2575 Sand Hill Road}, \city{Menlo Park, CA}, \postcode{94025}, \country{USA}}}

\affil[2]{\orgname{INFN}, \orgaddress{\street{16 Via Celoria}, \city{Milano}, \postcode{20133}, \country{Italy}}}


\abstract{Laser-driven Compton backscattering (CBS) has been proposed as method for controlling the intensity of colliding bunches in the FCC-ee so as to avoid the flip-flop instability caused by intensity asymmetry in colliding bunches. Laser-based collimation has also been proposed as an indestructible collimator for high-intensity electron beams. We have initiated a laboratory-based test program of these concepts with the E344 experiment at FACET-II. In this paper, we describe simulations of laser-beam interactions at FACET-II and the relevant scaling for FCC-ee. We also describe the experimental setup and diagnostics that will be used to make the measurements at FACET-II.}

\maketitle

\section{Introduction}\label{sec1}

Future colliders aim to maximize their luminosity in order to deliver on ambitious physics goals. In order to maximize luminosity, circular colliders operate with very large beam currents, and linear colliders operate with very small spot sizes. We encounter challenges as we push these machines to their luminosity limits. In the FCC-ee electron-positron collider~\cite{Benedikt:2928793}, which operates at high collision energy with intense beams, the beamstrahlung-induced flip-flop instability can lead to luminosity loss~\cite{shatilov1,shatilov2,PhysRevAccelBeams.27.121001}. The instability is driven by charge asymmetry in the colliding bunches, and can be mitigated by quickly removing charge from the more intense of the two bunches by Compton backscattering~\cite{zimmermann:ipac2022-wepost010,Drebot:2886031,drebot:ipac2025-mopm047}. In linear colliders, such as the ILC~\cite{Behnke:2013xla} and CLIC~\cite{CLIC:2012}, a considerable fraction of the beam delivery system (BDS) is devoted to collimating away high-amplitude halo particles. The collimators are physical devices that can be destroyed by the beam if the beam is mis-steered~\cite{Terui2024}. A laser-based collimation scheme has been proposed that would both reduce the length of the BDS and avoid the problem of collimator destruction~\cite{Zimmermann1999}. While both the laser control of bunch intensity and laser collimation of halo particles are promising topics, neither have been demonstrated in the lab. We have proposed the E344 experiment at FACET-II~\cite{Yakimenko2019} which leverages the E320 experimental infrastructure~\cite{Chen2022} to validate these concepts.

\section{Compton Back Scattering}\label{cbs}

During Compton backscattering, a high energy electron collides with a lower energy photon, and scatters the photon to modified energies and momenta \cite{Drebot:2886031}. The resulting electron energy after collision should be lower than its initial energy, thus providing a mechanism to remove unwanted electrons from the beam by pushing them outside the machine’s energy acceptance. To predict the number of Compton scattering events for the E320 experiment parameters, we compute the Compton cross section for the beam and laser parameters at FACET-II. The scattering parameter $x$ is defined as
\begin{equation}
    x=\dfrac{4 E_bE_\gamma}{m_e^2c^4}\cos^2\dfrac{\alpha_0}{2},
\end{equation}
where $E_b$ and $E_\gamma$ are the energies of the beam and photons respectively and $\alpha_0$ is the crossing angle between the incoming electron beam and laser. For the E320 experiment, $E_b=10 \text{ GeV}$, $\lambda_{\gamma}=800\nm$ and $\alpha_0=28.1^\circ$. The calculated scattering parameter is $x=0.223$.

The total Compton cross section $\sigma_0(x)$ is given by the formula \cite{Ginzburg:1982yr}:\begin{align*}
    \sigma_0(x)&=\dfrac{2\pi r_e^2}{x}\biggl([1-\dfrac{4}{x}-\dfrac{8}{x^2}]\ln(1+x)\\
    &+\dfrac{1}{2}[1-\dfrac{1}{(1+x)^2}]+\dfrac{8}{x}\biggl)\\
    &=550\text{ mbarn}.
\end{align*}
The maximum energy lost by scattered electrons is $E_{loss} = x(1+x)^{-1}E_b$. For the $10$ GeV electron beam, this amounts to $1.83$ GeV. The electron beam and laser pulse do not necessarily have the same size. The luminosity per bunch crossing for different-sized beams is given by \cite{drebot:ipac2025-mopm047, Herr:2003em}:  
\begin{gather*}
    \mathcal{L}=\dfrac{N_e N_\gamma}{2\pi \sqrt{\sigma_{e,x}^2+\sigma_{\gamma,x}^2 +(\sigma_{e,z}+\sigma_{\gamma,z})\tan\frac{\alpha_0}{2}}\sqrt{\sigma_{e,y}^2+\sigma_{\gamma,y}^2}}\\=5.3\times 10^{23}/\mu m^2,
\end{gather*}
where $N_e$ and $N_\gamma$ are the number of electrons and photons in the beam and laser pulses, respectively, $\alpha_0$ is the crossing angle, $\sigma_{e,x,y}$ and $\sigma_{\gamma,x,y}$ are the transverse size of the beam and laser pulses, respectively, and $\sigma_{e,z}$ and $\sigma_{\gamma,z}$ are the longitudinal sizes. Table~\ref{sizes} provides the beam parameters for the proposed experiment. Multiplying the cross section by the luminosity gives the total number of collisions for a single beam-laser interaction, with roughly 0.3\% of electrons scattering off a photon or $\mathcal{L}\sigma_0=3\times10^7$ events.

 \begin{table}[ht]
     \centering
     \begin{tabular}{lcccl}\toprule
          Species & $N$ & $\sigma_x$ ($\mu$m) & $\sigma_y$ ($\mu$m) &$\sigma_z$ \\ \midrule
          Electron & $10^{10}$ & 46 & 23  &20$\mu$m\\
          Photon & $4\times10^{17}$ & 10 & 10  &55 \text{fs}\\ \bottomrule
     \end{tabular}
     \caption{Electron and photon beam parameters for the proposed experiment at FACET-II.}
     \label{sizes}
 \end{table}
\vspace{-30pt}
\section{Modeling in CAIN and XSuite}
The beam dynamics were modeled through a combination of CAIN and Xsuite. CAIN is a FORTRAN Monte Carlo code that allows for the simulation of high energy photon, electron and positron interactions \cite{Yokoya:1985ab}. Xsuite is a tracking software that was used to model particles through the FACET-II lattice. CAIN was operated through the custom python interface Xcain that allows for streamlined communication between Xsuite and CAIN \cite{Hofer:xcain}. We assume that the electron and laser pulses are Gaussian distributions and note that the laser is round and smaller than the ellipsoidal electron beam, as shown in Table~\ref{sizes}.

       \begin{figure}[ht]
           \centering
           \includegraphics[width=0.8\linewidth]{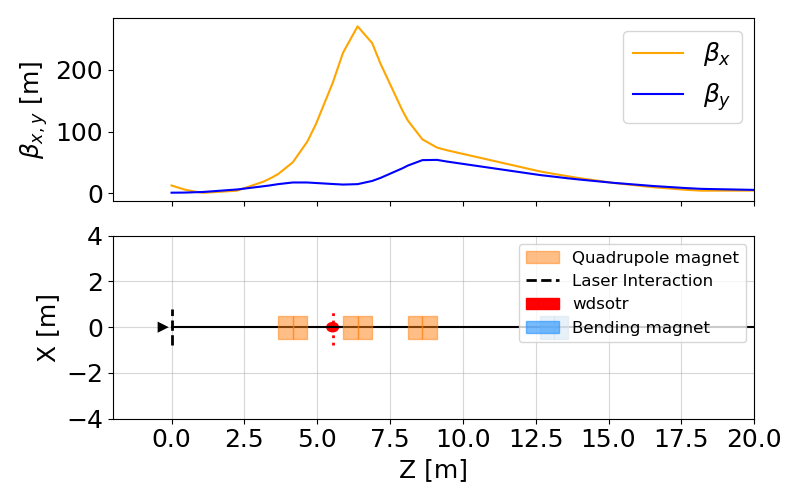}
           \caption{Top: $\beta_x$ and $\beta_y$ functions for this lattice section.
           Bottom: part of the final focusing section indicating the laser IP (black) and downstream OTR camera (red, ``wdsotr"). }
           \label{sector20}
       \end{figure}
\begin{figure}[ht]
    \centering
    \includegraphics[width=\linewidth]{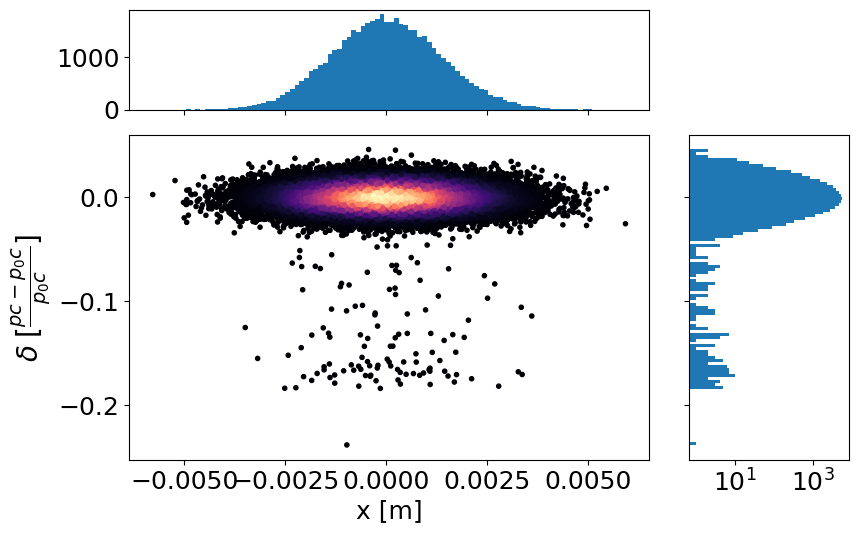}
    \caption{Energy and horizontal (x) position  after interaction with a $100\m\Joule$ laser after propagation through the FACET-II lattice. Energy is represented as a fractional deviation from the reference energy $p_0c=10\text{ GeV}$.}
    \label{simulation_delta_x}
\end{figure}

We represent the electron beam with $5\times 10^4$ macroparticles. We then initialize the laser element in CAIN, add it to the FACET beamline in Xsuite and track the macroparticles. The particles are initialized just upstream of the laser interaction point (IP) and then interact with the laser before propagating to a downstream optical transition radiation (OTR) screen. The beam betafunctions and a layout of the beamline are shown in Figure~\ref{sector20}. The location $z=0$ m corresponds to the IP and the red circle labled ``wdsotr" is the location of the OTR screen. Figure~\ref{simulation_delta_x}, shows the $x-\delta$ phase space at the location of the OTR screen. The energy distribution shows deviations of up to $\sim 20\% $, consistent with the estimate of 18\% calculated in Section II. The number of particles with an energy loss greater than than $3\%$ is about $2\times 10^7$ electrons, which is in good agreement with the predicted calculations and suggests that most Compton scattered electrons will be lost from the beam if they were subject to the FCC dynamic aperture~\cite{zimmermann:ipac2022-wepost010}.

\section{Experimental Configuration}

\begin{figure}[ht]
    \centering
    \includegraphics[width=\linewidth]{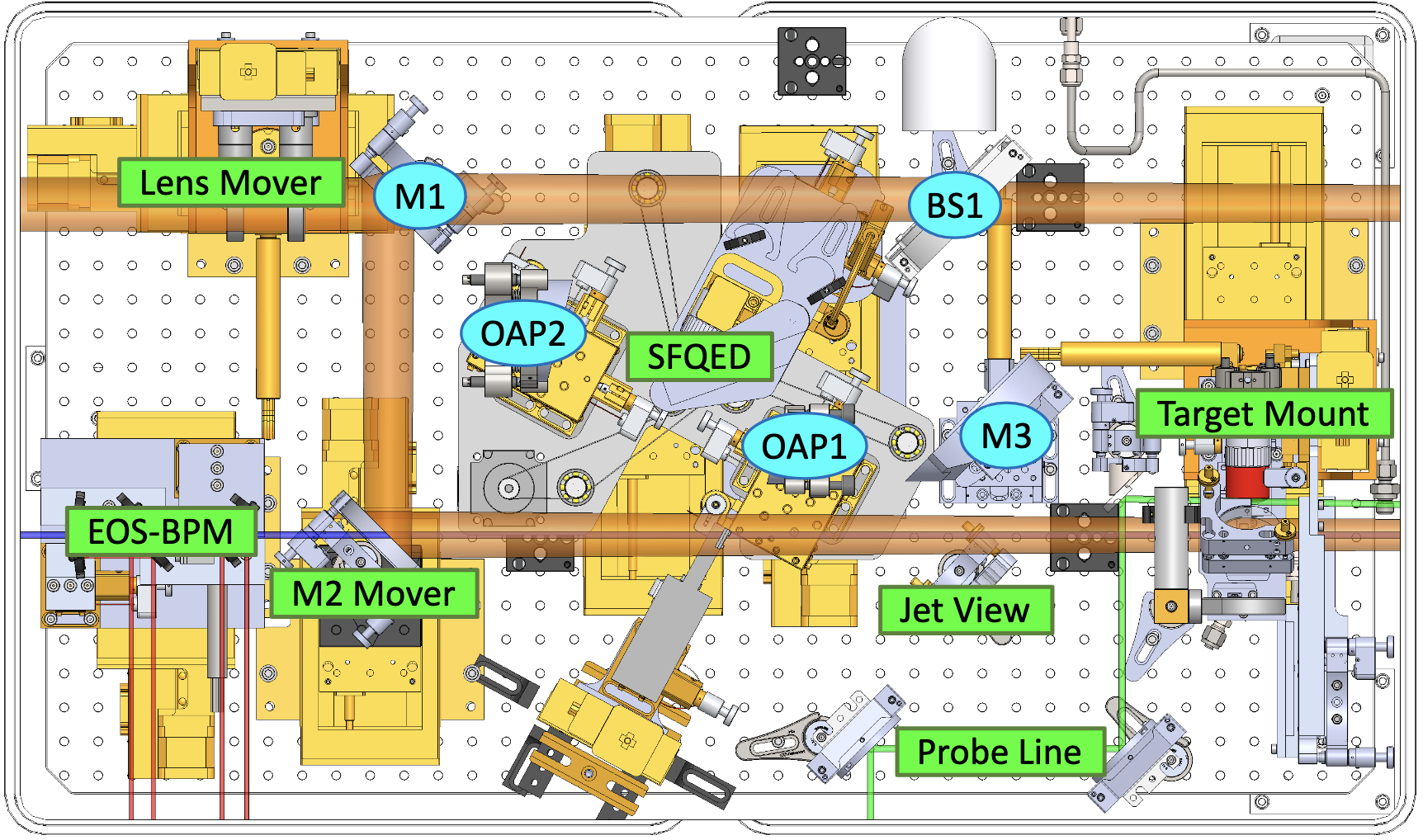}
    \caption{Schematic showing the E320 experimental layout inside the ``picnic basket" chamber at FACET-II. The electron beam enters from the left and passes by the electro-optical sampling (EOS) crystal which provides femtosecond-scale timing information for beam-laser synchronization. The main laser pulse enters from the top left and is directed by mirrors M1 and M2 onto the electron beam trajectory. The laser pulse is back-reflected onto the electron beam by the off-axis parabola (OAP) mirror. The OAP assembly is retractable and shown in the ``out" position in this schematic.}
    \label{picnic_basket}
\end{figure}

The beam-laser IP for the E320 experiment is shown in Figure~\ref{picnic_basket}. The bunch intensity control experiment will utilize the E320 experimental setup without modification. The maximum laser pulse energy on target is 0.5 J and the minimum laser pulse length is 55 fs, which yields an intense laser pulse suitable for nonlinear Compton scattering and strong-field QED experiments. The E344 experiment aims to operate in the linear Compton scattering regime, so we require only 0.1 mJ on target and a 100 fs-long laser pulse. The laser pulse energy is controlled by a waveplate and polarizer and the pulse length is controlled by adjusting the grating separation in the laser pulse compressor. The E320 experiment focuses the laser down to the smallest possible size and has demonstrated spots less than 2 $\mu$m wide. The E344 experiment will use the same off-axis parabola (OAP) focusing elements, but achieve larger spots by adding a defocusing term to an upstream deformable mirror.

For the laser collimation experiment, we propose to interact the beam with an annular laser, such that the laser only intercepts the beam halo particles. There are a number of ways to create annular laser pulses~\cite{Carbajo2025}. In previous work at FACET, we implemented high-order Bessel beams using diffractive optics~\cite{Gessner2016}. For the E344 laser collimation experiment, we will either replace the OAP focusing elements with diffractive optics, or add a spiral phase plate upstream of the OAP to achieve a Laguerre-Gauss mode at the focus.

\section{Diagnostics}
There are three primary diagnostics for the experiment. First, there is the electron energy spectrometer which consists of a vertical bending magnet and optical detector 10 meters downstream of the magnet. The dispersed electron beam produces Cherenkov radiation as it leaves the vacuum pipe just upstream of the beam dump. The Cherenkov light is intercepted by reflective wafers that direct the light to a camera which images the electron energy spectrum~\cite{Adli2015}. This imaging system is sensitive to sub-picocoulomb bunch charge and is able to detect the faint tail of decelerated electrons. In addition, a scintillating screen with enhanced charge sensitivity is also available to diagnose the tail of the electron energy distribution.

Second, a suite of gamma-ray diagnostics are used to measure the energy and profile of the CBS photon beam~\cite{Storey2024}. The gamma-rays are separated from the electron bunch as a result of the dipole spectrometer magnet and therefore can be imaged directly by scintillating screens. A filter wheel comprised of different materials attenuates the gamma-rays according to the radiation length and thickness of the material. We infer the energy spectrum of the gammas by measuring their attenuation by the filter wheel materials. 

        \begin{figure}[ht]
        \centering
        \includegraphics[width=0.8\linewidth]{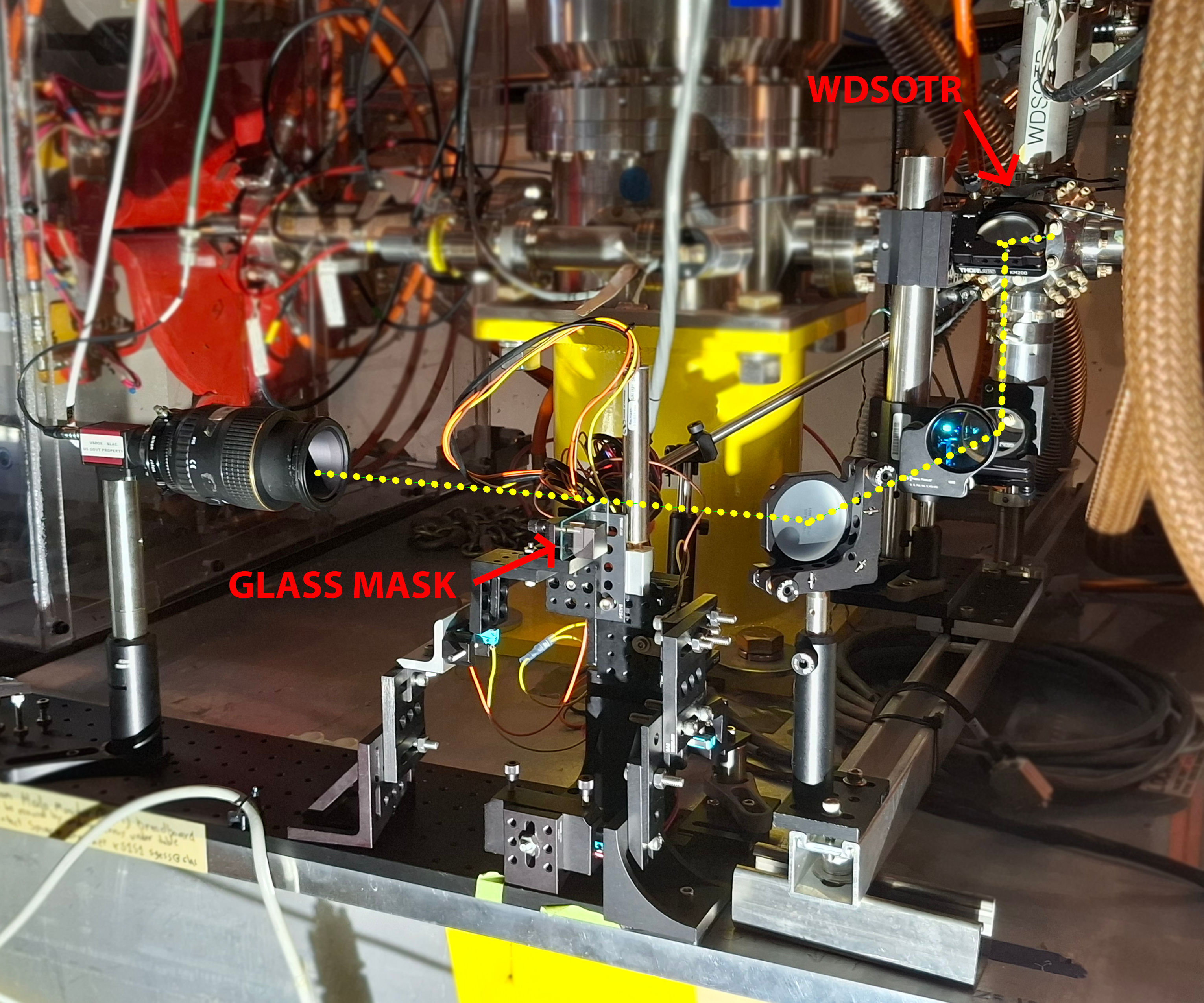}
        \caption{The BHM setup in the FACET-II tunnel at the WDSOTR location.}
        \label{imaging}
    \end{figure}

Third, we introduce a new diagnostic to FACET called the Beam Halo Monitor (BHM) at the downstream OTR location (``wdsotr" in Figure~\ref{sector20}). The BHM is an imaging station consisting of an OTR foil, a glass mask, lens, mirrors, and the camera. The diagnostic is modeled on the BHM setup used at AWAKE for proton beam-driven plasma wakefield acceleration experiments~\cite{turner2017}. The electron beam emits optical radiation as it passes through the foil, which normally operates as a transverse beam profile monitor. We are interested in observing solely the halo particles that interact with the annular laser in the collimation experiment, but the diagnostic is limited by the dynamic range of the camera. In order to image the beam halo, we re-image the light from the foil onto a glass slide with opaque dots of various diameters. The dots mask the core of the light while allowing the fainter halo light to pass through the glass slide. The mask diameters range from 0.1mm to 5mm, and the slide is attached to a motion control system that allow us to select and position the different masks. A camera with a lens objective images the light from the plane of the mask. Our version of the BHM has been implemented at FACET and is shown in Figure~\ref{imaging}.

\section{Conclusion}

The E344 experiment is prepared for beam time at FACET-II. The first set of experiments will demonstrate control of the electron-laser event rate which is required for bunch intensity management at FCC-ee. The laser bunch intensity control system for the FCC-ee will operate in a feedback mode and be used to adjust the bunch charge over many turns. FACET-II is a linear machine, so we will operate in a feedforward mode where measurements on a proceeding electron bunch determine the correction applied to a subsequent electron bunch. Our goal is to demonstrate intensity control at the part-per-mille level.

Our preparation for the collimation experiment with annular laser pulses requires rate estimates from theory and simulation. The CAIN code has an annular laser feature~\cite{Yokoya:1985ab}, but we have not successfully used this capability. We are involved in the development of alternative particle-in-cell (PIC) codes for beam-beam and beam-laser interactions, including WarpX~\cite{WarpX}. We plan to perform simulations of annular beam-laser interactions in WarpX and develop integrations with XSuite.

\bmhead{Acknowledgements}

This work is supported by SLAC National Accelerator Laboratory's LDRD program.


\section*{Declarations}

The simulation inputs, results and analysis can be found on github~\cite{munting_github}.

\bibliography{LCWS}

\end{document}